\documentclass{jfm}

\usepackage{graphicx}
\usepackage{comment} 
\usepackage{newtxtext}
\usepackage{newtxmath}
\usepackage{natbib}
\usepackage{hyperref}
\hypersetup{
	colorlinks = true,
	urlcolor   = blue,
	citecolor  = black,
}

\newcommand{\RomanNumeralCaps}[1]
\linenumbers

\usepackage[utf8]{inputenc} 				
\usepackage[main=english, ngerman]{babel}	
\usepackage[T1]{fontenc} 				    
\usepackage{textcomp} 					    
\usepackage{amsmath}					    
\usepackage{amssymb} 					    
\usepackage{mathtools}					    
\usepackage{cases}                          
\usepackage[noabbrev]{cleveref}			    
\usepackage[final]{microtype}			    
\usepackage{csquotes}					    
\usepackage{graphicx}                       
\usepackage{grffile}					    
\usepackage[subrefformat=parens]{subcaption}					    
\usepackage{booktabs}					    
\usepackage{xcolor}                         
\usepackage{tikz}                           
\usepackage{pgfplots}                       
\usepackage{placeins}                       
\usepackage{tabularx}                       
\usepackage{soul}
\usepackage[normalem]{ulem}

\pgfplotsset{compat=newest}
\usepgfplotslibrary{groupplots,dateplot,fillbetween}
\usetikzlibrary{patterns,shapes.arrows,calc,angles,quotes,babel,arrows,arrows.meta}
\pgfkeys{/pgf/number format/.cd,1000 sep={\,},min exponent for 1000 sep=4} 
\newlength\figureheight
\newlength\figurewidth

\newcommand{\walberla}{\textsc{waLBerla}}

\graphicspath{{./figures/}}

\title{Particle-resolved simulation of antidunes in free-surface flows}

\author{
	Christoph Schwarzmeier\aff{1}\corresp{\email{christoph.schwarzmeier@fau.de}}, 
	Christoph Rettinger\aff{1}, 
	Samuel Kemmler\aff{1}, 
	Jonas Plewinski \aff{1}, 
	Francisco N\'{u}\~{n}ez-Gonz\'{a}lez\aff{2},
	Harald K{\"{o}}stler\aff{1},
	Ulrich R{\"u}de\aff{1,3}
	\and Bernhard Vowinckel\aff{4}
}

\affiliation{
	\aff{1}Chair for System Simulation, Friedrich-Alexander-Universit{\"a}t Erlangen-N{\"u}rnberg, Cauerstra\ss e 11, 91058 Erlangen, Germany
	\aff{2}Department of Civil and Environmental Engineering, Universitat Politècnica de Catalunya • BarcelonaTech (UPC), Jordi Girona, 1-3, 08034 Barcelona, Spain. María Zambrano Fellow
    \aff{3}CERFACS, 42 Avenue Gaspard Coriolis, 31057 Toulouse Cedex 1, France
    \aff{4}Leichtwei\ss-Institute for Hydraulic Engineering and Water Resources, Technische Universit\"at Braunschweig, 38106 Braunschweig, Germany
}

\begin{document}
\maketitle

\begin{abstract}
The interaction of supercritical turbulent flows with granular sediment beds is challenging to study both experimentally and numerically; this challenging task has hampered the advances in understanding antidunes, the most characteristic bedform of supercritical flows.
This article presents the first numerical attempt to simulate upstream-migrating antidunes with geometrically resolved particles and a liquid--gas interface.
Our simulations provide data at a resolution higher than laboratory experiments, and they can therefore provide new insights into the mechanisms of antidune migration and contribute to a deeper understanding of the underlying physics.
To manage the simulations' computational costs and physical complexity, we employ the cumulant lattice Boltzmann method in conjunction with a discrete element method for particle interactions, as well as a volume of fluid scheme to track the deformable free surface of the fluid.
By reproducing two flow configurations of previous experiments (Pascal \emph{et al.}, Earth Surf.\ Proc.\ Land., vol.\ 46(9), 2021, 1750--1765), we demonstrate that our approach is robust and accurately predicts the antidunes' amplitude, wavelength, and celerity.
Furthermore, the simulated wall-shear stress, a key parameter governing sediment transport, is in excellent agreement with the experimental measurements.
The highly resolved data of fluid and particle motion from our simulation approach open new perspectives for detailed studies of morphodynamics in shallow supercritical flows.
\end{abstract}




\section{Introduction}\label{sec:introduction}

Antidunes are short-wave periodic disturbances that develop on the surface of loose granular beds in response to the interaction with supercritical and near-critical shallow, turbulent flows \citep{kennedy1963mechanics}, where criticality is defined by the Froude number.
These types of bedforms arise in fluvial, coastal, and submarine environments and are closely tied to the resulting flow resistance, turbulence, and sediment transport, which are quantities that are of great interest in geology as well as hydraulic and environmental engineering.
Among the bedforms at supercritical flow conditions, antidunes are the shortest in wavelength (on the order of the flow thickness), and they are also the only bedforms that can either migrate upstream, downstream, or even remain stationary.
This work deals with upstream migrating antidunes (UMAs).
The longitudinal profile of an UMA is symmetrical, exhibiting a sinusoidal shape that is mirrored by the undulating free water surface.
Consequently, UMAs are coupled to the fluid surface and are, thus, dependent on the presence of a free surface or an interface where stationary waves can develop.
The movement of UMAs in the direction opposite to the main flow is counterintuitive, and it is believed to result from the rhythmic aggradation of particles on the upstream flank of the bedform and erosion on the lee side.
However, little is known about the migration mechanism in connection to turbulence, bed morphology and sediment transport.
\par

Theoretical studies have made significant advances to understanding the characteristics of antidunes, by defining antidune instability regions \citep[e.g.,][]{kennedy1963mechanics, colombini2012three,bohorquez2019fascination} and identifying hydraulic constraints for antidune migration regimes \citep{nunez-gonzalez2011AnalysisAntiduneMigration}.
Unfortunately, due to the challenging supercritical flow conditions often associated with low submergences, experimental data sets to validate those theories are extremely sparse.
The limited number of systematic experimental data sets with antidune configurations is in part related to the inherent technical challenges to reproduce rapid flows over an erodible bed in laboratory flumes \citep{slootman2021supercritical}, as well as to the difficulties to perform non-intrusive measurements in flows which are generally unstable and shallow.
Numerical simulations of supercritical flows above an erodible bed, as a methodological alternative for the study of antidunes, are challenging because of the strongly varying water surface of rapid flows.
Existing simulations of supercritical bedforms have mainly considered the Saint--Venant shallow-water equations \citep[e.g.,][]{parker2000purely, balmforth2012cyclic} and the Reynolds-averaged Navier--Stokes approach to model cyclic steps \citep{vellinga2018morphodynamics} or downstream-migrating antidunes \citep{olsen2017numerical,olsen2022explaining}, but to the best of the authors' knowledge, simulations of UMAs have not been reported so far.
\par

Recently, \cite{kidanemariam2014direct} have successfully simulated downstream migrating ripples in subcritical flows, with particle-resolved direct numerical simulations in which the fluid--particle interaction is resolved by considering the motion of the flow and of each individual particle.
Owing to the subcritical flow conditions in their study, \cite{kidanemariam2014direct} could simplify the flow geometry by keeping the fluid depth constant.
In this work we propose to use particle-resolved simulations in conjunction with a deformable fluid surface, to simulate the formation and propagation of UMAs in supercritical flows.
We aim to numerically reproduce an experimental campaign recently reported by \cite{pascal2021VariabilityAntiduneMorphodynamics}, who managed to measure the propagation of UMAs with a high spatial and temporal resolution.
For this, we use the lattice Boltzmann method (LBM) to simulate the fluid hydrodynamics \citep{rettinger2022Coupling}  and extend it with a volume of fluid scheme \citep{schwarzmeier2023ComparisonFreesurfaceConservative} to track the strongly deformable free fluid surface in supercritical flows.
In turn, we couple the fluid hydrodynamics with the particle dynamics via the momentum-exchange method to simulate the antidune formation with high fidelity.
The parameter choices of \cite{pascal2021VariabilityAntiduneMorphodynamics}, with coarse sediment grains and low particle relative submergence, allow for a direct overlap of experimental conditions with simulations.
In this manner, the present work closes the gap between river morphodynamics’ experiments and numerical simulations, and demonstrates the capabilities offered by large-scale simulations of a mobile sediment bed comprising thousands of particles in unidirectional, supercritical turbulent flows, to study coupled bedform and free-surface interactions in great detail.
\par

\section{Numerical methods}\label{sec:nm}

All numerical simulations in this work were performed using the open-source multi-physics software framework \walberla{} \citep[][\url{https://www.walberla.net/}]{bauer2021WaLBerlaBlockstructuredHighperformance}.
In what follows, we briefly summarize the simulation's key features, namely the fluid--gas, the particle--particle, and the fluid--particle interactions.
\Cref{fig:method} presents a sketch of our numerical scheme.
\par

\begin{figure}
    \centering
    \begin{subfigure}[t]{0.36\textwidth}
        \vskip 0pt
        \begin{tikzpicture}[scale=0.7]
\tikzstyle{every node}=[font=\scriptsize]

\definecolor{interfaceLineBlue}{RGB}{31,119,180} 
\colorlet{liquidBlue}{interfaceLineBlue!50}
\colorlet{interfaceBlue}{interfaceLineBlue!25}

\definecolor{particleLineOrange}{RGB}{255,127,14} 
\colorlet{particleOrange}{particleLineOrange!33}

\fill[liquidBlue] (0,0) rectangle (6.5,4.5);

\fill[interfaceBlue] (0,4) rectangle ++(2,0.5);
\fill[interfaceBlue] (1.5,4.5) rectangle ++(3.5,0.5);
\fill[interfaceBlue] (4.5,4) rectangle ++(2,0.5);

\fill[particleOrange] (1,0.5) rectangle ++(1.5,2);
\fill[particleOrange] (0.5,1) rectangle ++(1,1);

\fill[particleOrange] (2.5,1.5) rectangle ++(0.5,1.5);
\fill[particleOrange] (3,1) rectangle ++(2.5,2.5);
\fill[particleOrange] (3,0.5) rectangle ++(2,0.5);

\draw[step=0.5,gray,very thin] (0,0) grid (6.5,5.5);

\begin{axis}%
[xmin=0,
xmax=12.925,
ymin=0,
ymax=10,
ticks=none,
axis lines=none,
clip=false,
scale only axis
]
\addplot[very thick, domain=0:10,samples=50,smooth,interfaceLineBlue] {6.15-0.5*cos(deg(pi*x*2)*5)};
\end{axis}

\coordinate[
] (xpi) at (1.7,1.6);
\coordinate[
] (xpj) at (4.15,2.1);
\node[fill=black, circle, inner sep=1.5] at (xpi) {};
\node[fill=black, circle, inner sep=1.5] at (xpj) {};
\draw[black, <->, >={Latex[scale=0.75]}] ($(xpi)+(0,-1)$) -- ($(xpi)+(0,+1)$) node[pos=0.3,left]{$d_{\text{p},j}$};
\draw[black, <->,  >={Latex[scale=0.75]}] ($(xpj)+(0,-1.6)$) -- ($(xpj)+(0,+1.6)$) node[pos=0.25,left]{$d_{\text{p},i}$};
\draw[particleLineOrange,very thick] (xpi) circle (1);
\draw[particleLineOrange,very thick] (xpj) circle (1.6);
\node[right] at (0.75,1.75) {$\rho_{\text{p}}$};
\node[right] at (4.75,1.25) {$\rho_{\text{p}}$};

\draw[-latex, thick] (xpi) -- ++(0.5,-0.5) node[pos=0.35,right]{$\boldsymbol{u}_{\text{p},j}$};
\draw[-latex, thick] (xpj) -- ++(-0.4,0.9) node[pos=0.5,left]{$\boldsymbol{u}_{\text{p},i}$};

\draw[-latex,thick] ($(xpi) + (150:1.3)$) arc (150:95:1.3);
\node[above] at ($(xpi) + (95:1.3)$) {$\boldsymbol{\omega}_{\text{p},j}$};
\draw[-latex,thick] ($(xpj) + (-10:1.9)$) arc (-10:35:1.9);
\node[above] at ($(xpj) + (35:1.9)$) {$\boldsymbol{\omega}_{\text{p},i}$};

\node at (3.25,5.25) {$p_{\text{g}}$\strut};
\node[right,interfaceLineBlue] at (4.5,4.75) {$p_{\text{l},\text{L}}(\boldsymbol{x},t)$\strut};
\node[right,interfaceLineBlue] at (0.5,4.75) {$\kappa_{\text{l}}(\boldsymbol{x},t)$\strut};

\node[right] at (0,0.25) {$\rho_{\text{l}}(\boldsymbol{x},t) \quad \boldsymbol{u}_{\text{l}}(\boldsymbol{x},t) 
\quad \nu_{\text{l}}$\strut};

\draw[-, very thick, interfaceLineBlue] (0,-0.4)--(0.35,-0.4);
\node[right] at (0.35,-0.4) {liquid--gas interface\strut};
\draw[gray, very thin, fill=interfaceBlue] (0,-1.075) rectangle (0.35,-0.725);
\node[right] at (0.35,-0.9) {interface cell\strut};
\draw[gray, very thin, fill=liquidBlue] (0,-1.575) rectangle (0.35,-1.225);
\node[right] at (0.35,-1.4) {liquid cell\strut};

\draw[-, very thick, particleLineOrange] (4.5,-0.4)--(4.85,-0.4);
\node[right] at (4.85,-0.4) {particle\strut};
\draw[gray, very thin, fill=particleOrange] (4.5,-1.075) rectangle (4.85,-0.725);
\node[right] at (4.85,-0.9) {solid cell\strut};
\draw[gray, very thin, fill=white] (4.5,-1.575) rectangle (4.85,-1.225);
\node[right] at (4.85,-1.4){gas cell\strut};

\node at (0.25,5.8) {\normalsize{(a)}};
\end{tikzpicture}%
        \phantomcaption\label{fig:method}%
    \end{subfigure}
    \hfill%
    \setlength{\figurewidth}{0.475\textwidth}%
    \begin{subfigure}[t]{0.625\textwidth}
        \vskip 0pt
        \input{figures/setup-simulation.tex}%
        \phantomcaption\label{fig:setup}%
    \end{subfigure}
    \caption{
        Schematic representation of the simulations, coupling the liquid (l) with a particle (p) and gas (g) phase~\subref{fig:method}, and the simulation setup in its initial condition~\subref{fig:setup}.
        \Cref{fig:setup} is a zoom into the computational domain covering only 16\,\% of its streamwise extent.
    }
\end{figure}

\subsection{Free-surface lattice Boltzmann method}\label{subsec:lbm}

The LBM \citep{kruger2017LatticeBoltzmannMethod} discretizes the Boltzmann equation from kinetic theory and describes the evolution of particle distribution functions (PDFs) on a uniform computational grid to simulate a fluid flow.
Macroscopically, the LBM approximates the Navier--Stokes equations with second-order accuracy in space $\boldsymbol{x}$ and time $t$.
The cell-local fluid density $\rho_{\text{l}}(\boldsymbol{x},t)$ and velocity $\boldsymbol{u}_{\text{l}}(\boldsymbol{x},t)$ are given by the zeroth- and first-order moments of the cell's PDFs.
In the remainder of this article, all quantities denoted with the superscript \textit{LU} are specified in a normalized LBM unit system.
We set the cell size $\Delta x^{\text{LU}} = 1$, time step size $\Delta t^{\text{LU}} = 1$, and the reference density of the fluid $\rho^{\text{LU}}_{\text{l},\text{ref}}=1$.
We use the D3Q27 cumulant collision model proposed by \citet{geier2015CumulantLatticeBoltzmann} and set all relaxation rates to unity, except for $\omega^{\text{LU}}$ that defines the kinematic viscosity of the fluid $\nu_{\text{l}}^{\text{LU}} = 1/3 (1/\omega^{\text{LU}} - \Delta t^{\text{LU}}/2)$.
We model no-slip boundary conditions at solid obstacles using the bounce-back rule \citep{kruger2017LatticeBoltzmannMethod}.
\par

The \walberla{} framework includes an implementation of the free-surface lattice Boltzmann method \citep[FSLBM, ][]{schwarzmeier2023ComparisonFreesurfaceConservative}.
The FSLBM is based on a \textit{volume of fluid} scheme and simulates the interface between two immiscible fluids.
The model assumes that the liquid fluid (higher density) governs the flow dynamics of the system so that the fluid flow in the gaseous fluid (lower density) is neglected.
A free-surface boundary condition captures the motion of the liquid--gas interface.
The boundary condition incorporates the gas pressure $p_{\text{g}}$ and the Laplace pressure $p_{\text{l},\text{L}}(\boldsymbol{x},t)=2\sigma_{\text{l}}\kappa_{\text{l}}(\boldsymbol{x},t)$, where $\sigma_{\text{l}}$ is the surface tension of the liquid, and $\kappa_{\text{l}}(\boldsymbol{x},t)$ is the local curvature of the free surface.
\par

\subsection{Particle--particle and fluid--particle coupling}\label{subsec:dem}

The temporal evolution of the translational and rotational velocity, $\boldsymbol{u}_{\text{p},i}$ and $\boldsymbol{\omega}_{\text{p},i}$, of an individual rigid particle $i$ is described by the Newton--Euler equations for rigid-body motion.
The total force and torque acting on a particle comprises contributions from inter-particle collisions, hydrodynamic interactions, and an external force attributed to the gravitational acceleration $\boldsymbol{g}$.
The inter-particle collisions are modeled via the discrete element method (DEM) specifically designed for a four-way coupled lattice Boltzmann method of particle-laden flows. The model is fully parameterized by the dry coefficient of restitution of $e_{\text{dry}} = 0.97$, the collision time $T_{\text{c}}^{\text{LU}} = 4 d_{50}^{\text{LU}} \Delta t^{\text{LU}} /\Delta x^{\text{LU}}$, the Poisson's ratio $\nu_{\text{p}} = 0.22$, and the friction coefficient $\mu_{\text{p}}=0.5$ to simulate silica beads.
Details on the particle--particle and fluid--particle coupling can be found in \citet{rettinger2022Coupling}.
\par

For fluid--particle interactions, we employ a geometrically resolved coupling between the fluid and the particulate phase.
We impose no-slip boundary conditions for the fluid along the particle surface and compute the hydrodynamic forces and torques acting on the particles directly.
This geometrically resolved coupling is achieved by the LBM-specific momentum-exchange method (MEM).
The MEM requires an explicit mapping of the particles onto the underlying computational grid \citep{aidun1998,wen2014}.
Following \citet{rettinger2022Coupling}, we additionally augment the DEM with lubrication corrections in the normal and tangential directions to model these strong subgrid-scale hydrodynamic forces for situations where gaps between particles become too small to be properly resolved by the LBM mesh.
\par

\section{Simulation setup}\label{sec:setup}

\subsection{Physical scenario and simulation parameterization}\label{subsec:parameters}
The simulations carried out for this article aim to reproduce the laboratory experiments of \citet{pascal2021VariabilityAntiduneMorphodynamics}, who performed their experiments with water and a sediment bed of natural gravel.
The gravel particles had a density of $\rho_{\text{p}} = 2550$\,kg/m\textsuperscript{3} and characteristic sieving diameters of $d_{16}=2.5$\,mm, $d_{50}=2.9$\,mm, and $d_{84}=3.3$\,mm, where the subscripts denote the respective percentile of the particle size distribution.
The mean flow velocity $U_{x,\text{l}} = q_{\text{w}} / h_{0}$ is computed from the measured flow discharge in the surface flow layer $q_{\text{w}}$ and reference flow depth $h_{0}$ (\cref{fig:setup}).
Three dimensionless numbers specify the flow conditions:
the Reynolds number $\mathit{Re} = U_{x,\text{l}} h_{0} / \nu_{\text{l}}$, the Froude number $\mathit{Fr} = U_{x,\text{l}}/\sqrt{g h_{0}}$, and the Weber number $\mathit{We} = \rho_{\text{l}} U_{x,\text{l}}^{2} h_{0} / \sigma_{\text{l}}$, where  $\nu_{\text{l}} = 1 \cdot 10^{-6}$\,m\textsuperscript{2}/s and $\sigma_{\text{l}} = 7.2 \cdot 10^{-2}$\,kg/s\textsuperscript{2} are the kinematic viscosity and surface tension of water, respectively.
The values of these numbers are listed in \cref{tab:simulation-conditions}, where we refer to the simulation setups with the same labels E1 and E4, as chosen by \citet{pascal2021VariabilityAntiduneMorphodynamics}.
\par

\begin{table}
  \centering
  \begin{tabular}{
                    >{\raggedright}m{0.075\textwidth}
		    	    >{\centering\arraybackslash}m{0.075\textwidth}
			        >{\centering\arraybackslash}m{0.075\textwidth}
			        >{\centering\arraybackslash}m{0.075\textwidth}
			        >{\centering\arraybackslash}m{0.125\textwidth}
			        >{\centering\arraybackslash}m{0.125\textwidth}
			        >{\centering\arraybackslash}m{0.1\textwidth}
			        >{\centering\arraybackslash}m{0.1\textwidth}
                    >{\centering\arraybackslash}m{0.1\textwidth}}
    Label & $\mathit{Re}$ & $\mathit{Fr}$ & $\mathit{We}$ & $\Delta x$\,/\,m & $\Delta t$\,/\,s & $\omega^{\text{LU}}$ & $U^{\text{LU}}_{x,\text{l}}$ & $h_{\text{0}}$ \\
    \midrule
    E1 & $3100$ & $1.31$ & $15.62$ & $2.5 \cdot 10^{-4}$ & $1.38 \cdot 10^{-5}$ & $1.997$ & $0.02$ & $2.97 d_{50}$ \\
    E4 & $4800$ & $1.45$ & $30.25$ & $2.5 \cdot 10^{-4}$ & $1.08 \cdot 10^{-5}$ & $1.998$ & $0.02$ & $3.59 d_{50}$
  \end{tabular}
  \caption{\label{tab:simulation-conditions}
    Summary of the main simulation conditions.
    The labels are chosen as in the reference experiments from \citet{pascal2021VariabilityAntiduneMorphodynamics}.
  }
\end{table}

The simulation domain was discretized with a mesh size $\Delta x = 2.5 \cdot 10^{-4}$\,m on a uniform computational grid of size $L^{\text{LU}}_x \times L^{\text{LU}}_y \times L^{\text{LU}}_z = 3200 \times 60 \times 160$ cells, corresponding to $0.8\,\text{m} \times 0.015\,\text{m} \times 0.04\,\text{m}$ in the streamwise, spanwise and vertical direction, respectively.
We have used periodic boundary conditions in the $x$- and $y$-directions.
The boundaries in the $z$-direction and on the particle surface were modeled with a no-slip condition.
The sediment bed consisted of $6528$ spherical particles with a mean diameter of $d^{\text{LU}}_{50} = 11.6$, where the actual diameters were sampled from a uniform distribution of the interval $[0.9,1.1] d_{50}$.
We have generated the initially flat particle bed of height $h_{\text{b}} = 4.14 d_{50}$ using a rigid body dynamics simulation for a sedimentation process in dry conditions.
To this end, particles were dropped onto a single layer of fixed particles that were positioned on a horizontal hexagonal grid with small random perturbations in their $z$-positions.
We fix all particles with a vertical center coordinate $z_{\text{p},i}<h_{\text{b,f}} = 1.5 d_{50}$ to generate a rough bottom.
We have carried out sensitivity studies to verify that the chosen number of particles and $h_{\text{b,f}}$ were sufficient to exclude finite-size effects of the domain extent.
\par

The liquid was initialized with a height of $h_{\text{b}}+h_{0}$, where $h_{0}$ represents the initial height of the liquid on top of the sediment bed, and with a parabolic velocity profile on top of the bed.
The hydrostatic density was initialized, such that the liquid pressure was equal to the constant atmospheric gas pressure at the free surface.
The flow was driven in $x$-direction via an externally imposed body force $F_x$, which was controlled during the simulation to maintain the desired bulk flow rate.
The mean fluid velocity was set to $U^{\text{LU}}_{x,\text{l}} = 0.02$, leading to the time step sizes $\Delta t$ specified in \cref{tab:simulation-conditions}.
\par

\subsection{Evaluation quantities and grid resolution study}\label{subsec:evaluation}

The temporal and spatial variability of the bed topography is one of the key quantities to characterize the evolution of antidunes.
We follow the approach of \citet{kidanemariam2014direct} and compute the $y$-averaged solid-volume fraction $\phi_{\text{b}}(x,z,t)$ for each computational cell in the $x$-, $z$-plane.
The bed height $h_{\text{b}}(x,t)$ is then given as the linearly interpolated $z$-value, for which $\phi_{\text{b}}(x,z,t) = 0.3$.
The bedload transport rate $q_{\text{b}}(t)$ is computed via
\begin{equation}\label{eq:bedload-transport-rate}
    q_{\text{b}}(t)=\frac{\sum_{i \in \text{P}} u_{x,\text{p},i}(t) V_{\text{p},i}}{L_{x}L_{y}},
\end{equation}
where P is the set of all particles and $u_{x,\text{p},i}(t)$ is the streamwise particle velocity of particle $i$.
\par

\begin{figure}
    \centering
	\setlength{\figureheight}{0.3\textwidth}
	\setlength{\figurewidth}{\textwidth}
	\input{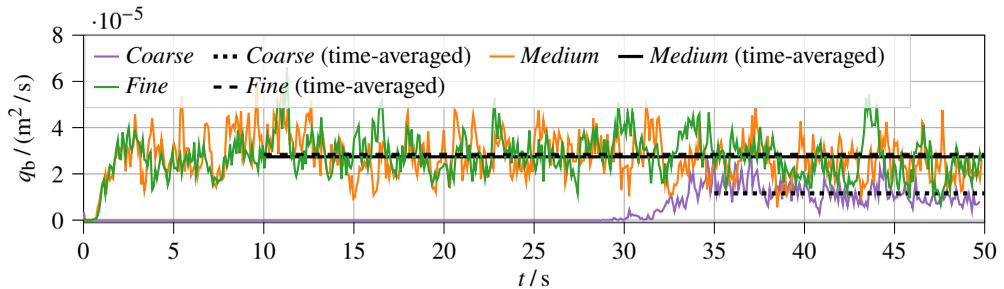}%
    \caption{\label{fig:resolution-study}
        Simulated bedload transport rate $q_{\text{b}}(t)$ with different computational grid resolutions using a shortened simulation domain.
        The time-averages were computed only in the time-ranges conforming with the lengths of the plotted lines to ensure that the fluid flow was adequately developed.
    }
\end{figure}

We carried out a grid sensitivity study to verify that our results are independent of numerical parameters.
To this end, we focused on the more turbulent flow conditions of E4 (see \cref{tab:simulation-conditions}).
In addition to the reference case from \cref{subsec:parameters} with  $\Delta x_{50,\textit{Medium}} = 2.5 \cdot 10^{-4}$\,m as a  medium resolution, we have tested two additional cases with a two-times coarser ($\Delta x_{50,\textit{Coarse}} = 5 \cdot 10^{-4}$\,m) and finer resolution ($\Delta x_{50,\textit{Fine}} = 1.25 \cdot 10^{-4}$\,m).
These computational grid resolutions lead to $d^{\text{LU}}_{50,\textit{Coarse}} = 5.8$, $d^{\text{LU}}_{50,\textit{Medium}} = 11.6$, and $d^{\text{LU}}_{50,\textit{Fine}} = 23.2$.
For the grid sensitivity study, we shortened the simulation domain by a factor of $10$ in the $x$-direction to keep the computational cost manageable.
We have scaled the time step size $\Delta t$ proportionally to $\Delta x$ in our study with reference $\Delta t_{50,\textit{Medium}} = 1.08 \cdot 10^{-5}$\,s, as listed in \cref{tab:simulation-conditions}.
\Cref{fig:resolution-study} compares the bedload transport rates~\eqref{eq:bedload-transport-rate} for the different grid resolutions.
From this study, we conclude that the grid \textit{Medium} is sufficiently independent of the grid resolution and can be used for further analysis.
\par

\section{Results}\label{sec:results}

\begin{figure}
	\setlength{\figurewidth}{1\textwidth}	
	\centering
	\input{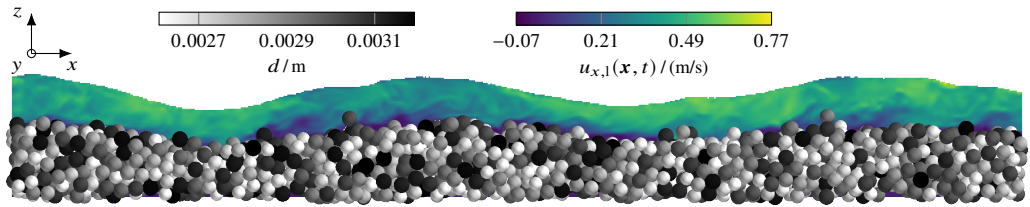}%
	\caption{\label{fig:velocity-field}
	    Visualization of the simulated velocity in streamwise direction $u_{x,\text{l}}(\boldsymbol{x},t)$.
	    The undulation of the sediment bed and free surface of the liquid are in phase, conforming to the definition of antidunes.
        The figure is a zoom into the computational domain covering only $25$\,\% of its streamwise extent.
	}
\end{figure}

Our particle-resolved simulations provide the position of each particle as well as the location of the water surface in space and time, and full fluid information (\cref{fig:velocity-field}).
Bed elevation perturbations arose spontaneously from the initial flat bed right from the start of the simulations.
From then on, a regular longitudinal pattern of quasi-periodic bedforms, with gentle slopes up- and downstream of the crests, prevailed throughout the simulation time.
The bed patterns were approximately in phase with respect to the free surface and they slowly migrated upstream (see animations in supplementary material).
Therefore, these bedforms can be unequivocally classified as UMAs that compare well to the observations of the experiments of \cite{pascal2021VariabilityAntiduneMorphodynamics}.
A data set with such a high resolution as the one generated here numerically, has so far not been available from physical experiments on movable beds under supercritical flows.
\par

\subsection{Bed elevation perturbations}\label{subsec:bed-elevation}

\begin{figure}
	\centering
	\setlength{\figurewidth}{0.42\textwidth}
	\begin{minipage}{0.9\textwidth}
		\begin{subfigure}[t]{0.49\textwidth}
            \setlength{\figureheight}{0.7\textwidth}%
	        \vskip 0pt
\begin{tikzpicture}

\definecolor{darkgray176}{RGB}{176,176,176}

\begin{axis}[
xtick={0,0.25,...,0.75},
colormap/viridis,
height=\figureheight,
point meta max=0.005,
point meta min=-0.005,
tick align=outside,
tick pos=left,
width=\figurewidth,
x grid style={darkgray176},
xlabel={},
xticklabels={,,},
xmin=0, xmax=0.75,
xtick style={color=black},
y grid style={darkgray176},
ylabel={$t$\,/\,s},
ymin=1200, ymax=1245,
ytick style={color=black},
ytick={1200,1215,...,1245},
clip mode=individual,
yticklabel style={text width=20 pt,align=right}
]
\addplot graphics [includegraphics cmd=\pgfimage,xmin=0, xmax=0.75, ymin=1200, ymax=1245] {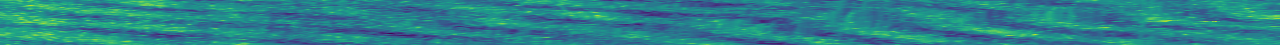};

\node[anchor=south west,above right,yshift=0.1em,inner sep=0] at (axis description cs:0,1)  {(a) E1 experiment};
\end{axis}

\end{tikzpicture}%
	        \phantomcaption\label{fig:bed-elevation-experiment-e1}%
		\end{subfigure}
		\hfill%
  		\begin{subfigure}[t]{0.49\textwidth}
            \setlength{\figureheight}{0.7\textwidth}%
	        \vskip 0pt
\begin{tikzpicture}

\definecolor{darkgray176}{RGB}{176,176,176}

\begin{axis}[
xtick={0,0.25,...,0.75},
colormap/viridis,
height=\figureheight,
point meta max=0.005,
point meta min=-0.005,
tick align=outside,
tick pos=left,
width=\figurewidth,
x grid style={darkgray176},
xlabel={},
xticklabels={,,},
xmin=0, xmax=0.75,
xtick style={color=black},
y grid style={darkgray176},
ylabel={$t$\,/\,s},
ymin=1200, ymax=1245,
ytick style={color=black},
ytick={1200,1215,...,1245},
clip mode=individual,
yticklabel style={text width=20 pt,align=right}
]
\addplot graphics [includegraphics cmd=\pgfimage,xmin=0, xmax=0.75, ymin=1200, ymax=1245] {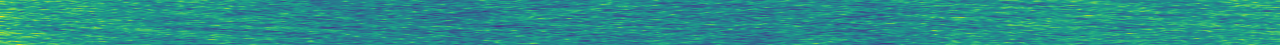};

\node[anchor=south west,above right,yshift=0.1em,inner sep=0] at (axis description cs:0,1) {(c) E4 experiment};
\end{axis}

\end{tikzpicture}%
	        \phantomcaption\label{fig:bed-elevation-experiment-e4}%
		\end{subfigure}

		\begin{subfigure}[t]{0.49\textwidth}
            \setlength{\figureheight}{0.98\textwidth}%
	        \vskip 0pt
\begin{tikzpicture}

\definecolor{darkgray176}{RGB}{176,176,176}

\begin{axis}[
xtick={0,0.25,...,0.75},
ytick={0,15,...,75},
colormap/viridis,
height=\figureheight,
point meta max=0.005,
point meta min=-0.005,
tick align=outside,
tick pos=left,
width=\figurewidth,
x grid style={darkgray176},
xlabel={$x$\,/\,m},
xlabel style={yshift=0.25em},
xmin=0, xmax=0.75,
xtick style={color=black},
y grid style={darkgray176},
ylabel={$t$\,/\,s},
ymin=0, ymax=75,
ytick style={color=black},
clip mode=individual,
yticklabel style={text width=20 pt,align=right}
]
\addplot graphics [includegraphics cmd=\pgfimage,xmin=0, xmax=0.75, ymin=0, ymax=75] {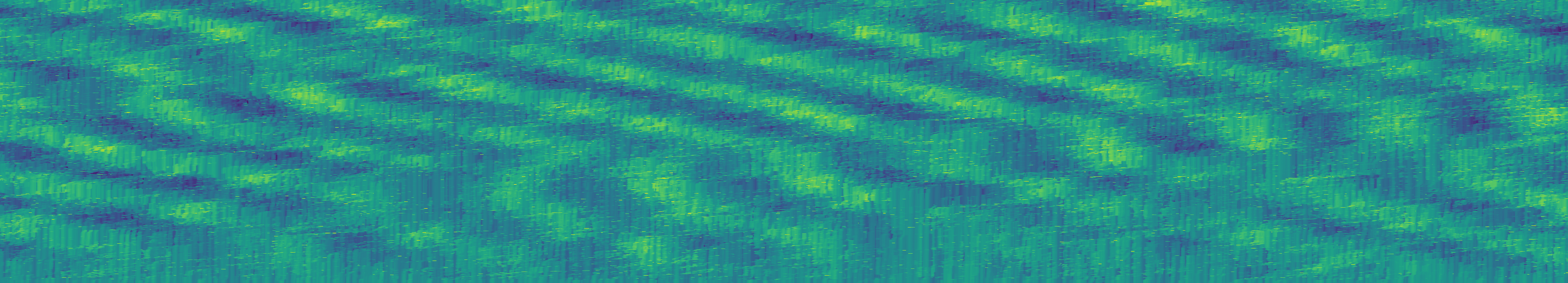};
\addplot [thick, dashed, black]
table {%
0 30
75 30
};
\node[anchor=south west,above right,yshift=0.1em,inner sep=0] at (axis description cs:0,1) {(b) E1 simulation};
\end{axis}

\end{tikzpicture}%
	        \phantomcaption\label{fig:bed-elevation-simulation-e1}%
            \vspace{-1.45\baselineskip}
		\end{subfigure}
		\hfill%
		\begin{subfigure}[t]{0.49\textwidth}
            \setlength{\figureheight}{0.98\textwidth}%
	        \vskip 0pt
\begin{tikzpicture}

\definecolor{darkgray176}{RGB}{176,176,176}

\begin{axis}[
xtick={0,0.25,...,0.75},
ytick={0,15,...,75},
colormap/viridis,
height=\figureheight,
point meta max=0.005,
point meta min=-0.005,
tick align=outside,
tick pos=left,
width=\figurewidth,
x grid style={darkgray176},
xlabel={$x$\,/\,m},
xlabel style={yshift=0.25em},
xmin=0, xmax=0.75,
xtick style={color=black},
y grid style={darkgray176},
ylabel={$t$\,/\,s},
ymin=0, ymax=75,
ytick style={color=black},
clip mode=individual,
yticklabel style={text width=20 pt,align=right}
]
\addplot graphics [includegraphics cmd=\pgfimage,xmin=0, xmax=0.75, ymin=0, ymax=75] {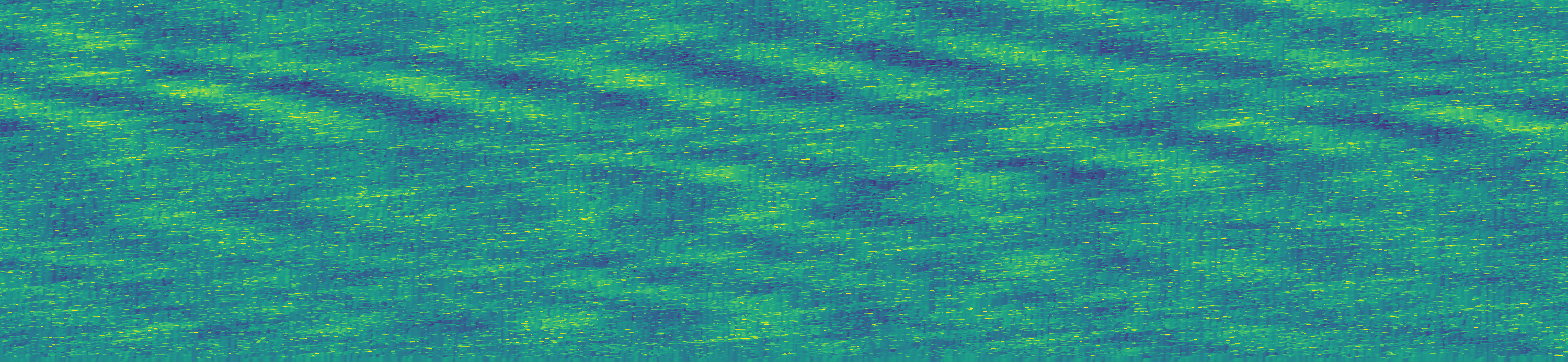};

\addplot [thick, dashed, black]
table {%
0 30
75 30
};

\node[anchor=south west,above right,yshift=0.1em,inner sep=0] at (axis description cs:0,1) {(d) E4 simulation};
\end{axis}

\end{tikzpicture}%
	        \phantomcaption\label{fig:bed-elevation-simulation-e4}%
            \vspace{-1.45\baselineskip}
		\end{subfigure}
	\end{minipage}
	\begin{minipage}{0.09\textwidth}
		\begin{tikzpicture}
			\pgfplotscolorbardrawstandalone[ 
    			colormap/viridis,
    			point meta max=0.005,
    			point meta min=-0.005,
    			colorbar style={
    				ylabel={$h_{\text{b}}$\,/\,m},
    				width=0.2\textwidth,
    				height=3\textwidth,
    				every y tick scale label/.style={
    					at={(yticklabel* cs:1.075,0)},
    					anchor=near yticklabel
    				},
    			}
			]
		\end{tikzpicture}
		\vspace{1.5\baselineskip}
	\end{minipage}
	\caption{\label{fig:bed-elevation}
	    Sediment bed elevation $h_{\text{b}}(x,t)$ for E1 (left) and E4 (right): \subref{fig:bed-elevation-experiment-e1} and \subref{fig:bed-elevation-simulation-e1} data from \citet{pascal2021VariabilityAntiduneMorphodynamics}, and \subref{fig:bed-elevation-experiment-e4} and \subref{fig:bed-elevation-simulation-e4} data from the numerical simulations.
        We consider the system fully developed after $t=30$\,s, as illustrated by the dashed black line.
	}
\end{figure}

Space--time evolution diagrams of the bed surface, for the simulated and the experimental bedforms, are plotted in \cref{fig:bed-elevation}.
Such diagrams result from laterally-averaged bed surface profiles at every 1000 time steps of the simulation, and from footage recorded from the transparent channel walls in the experiments of \cite{pascal2021VariabilityAntiduneMorphodynamics}.
From the original experimental data,  we have selected the intervals corresponding to the lower left parts of figures 3a and 3d in this reference, to reproduce a part of their bed elevation plots to compare with the scales of our numerical data.
The alternating blue and yellow (dark and light) diagonal strips denote the troughs and crests of the bedforms, respectively.
The experimental and numerical diagrams with the evolution of the bottom elevation in the spatial--temporal domain are in good qualitative agreement.
A similar upstream migration trend of the bedforms is clearly visible due to the negative slope of the strips.
In some regions the strips bend, indicating acceleration/deceleration, or even stationarity of the bedforms.
Particularly for run E1, these regions are related to local perturbations that migrate downstream. Note that similar perturbations can be identified in both experimental and simulation plots.
Overall, the simulated bedforms appear stable and compare well to their experimental counterparts.
\par

As pointed out by \cite{pascal2021VariabilityAntiduneMorphodynamics}, the definition of a dominant bedform migration speed seems inappropriate given the nonuniformity of the bedform celerities observed.
The same reasoning is valid for the amplitude and the wavelength of the antidunes.
In our simulations, the amplitude of the simulated bed undulations ranged approximately from $1$ to $3$ times the particle median grain size $d_{50}$, and the wavelength ranged from about $10$ to $15$ times the water depth $h_0$.
These values are in good quantitative agreement with the experimental data, where the dune amplitudes ranged from one particle median grain size to the mean flow depth ($\approx 3d_{\text{50}}$), and typical wavelengths varied from $0.05$ to $0.15$\,m ($\approx 6h_{0}$ to $15h_{0}$).
\par

\begin{figure}
	\centering
	\setlength{\figureheight}{0.3\textwidth}
	\setlength{\figurewidth}{0.365\textwidth}
	\begin{subfigure}[t]{0.495\textwidth}
        \vskip 0pt
		\begin{tikzpicture}

\definecolor{darkgray176}{RGB}{176,176,176}
\definecolor{lightgray204}{RGB}{204,204,204}
\definecolor{darkorange25512714}{RGB}{255,127,14}

\begin{axis}[
axis on top,
colorbar,
colorbar style={
	ylabel={PSD},
	width=0.05\figurewidth,
	at={(1.1,0)},
	anchor=south west,
	ytick={0,0.004,...,0.016},
	every y tick scale label/.style={
		at={(yticklabel* cs:1.2,0)},
		anchor=near yticklabel
	},
},
point meta min=0,
point meta max=0.016,
colormap/viridis,
height=\figureheight,
tick align=outside,
tick pos=left,
width=\figurewidth,
x grid style={darkgray176},
xlabel={$T$\,/\,s},
xlabel style={yshift=0.25em},
xmin=5, xmax=70,
xtick style={color=black},
y grid style={darkgray176},
ylabel={$\lambda$\,/\,m},
scaled y ticks=false,
yticklabel style={
	/pgf/number format/fixed,
	/pgf/number format/precision=3
},
ymin=0.05, ymax=0.2,
ytick style={color=black},
xtick={5,25,...,70},
clip mode=individual,
legend cell align={left},
legend style={
    legend columns = 2,
	fill opacity=0.8,
	draw opacity=1,
	text opacity=1,
	at={(0.999,0.999)},
	anchor=north east,
	draw=lightgray204,
    font=\scriptsize,
    legend image post style={scale=0.55}
},
]
\addlegendimage{line legend,darkorange25512714,very thick}
\addlegendentry{Simulation}

\addplot graphics[xmin=5,xmax=70,ymin=0.05,ymax=0.2]{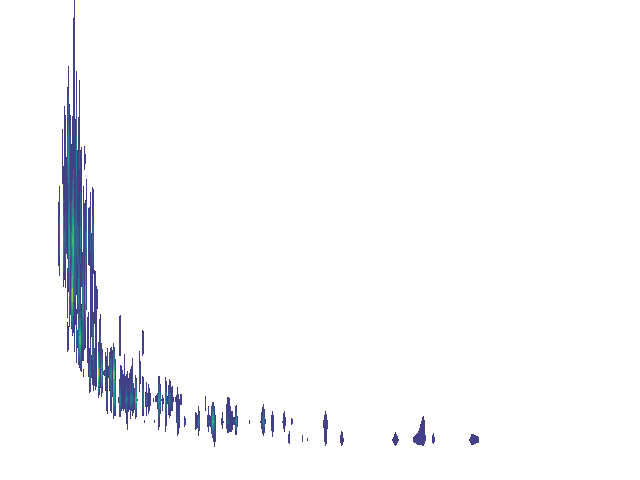};
\addplot graphics[xmin=5,xmax=70,ymin=0.05,ymax=0.2]{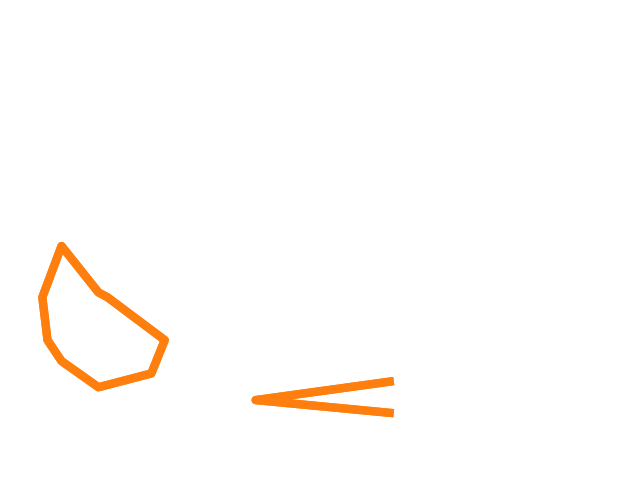};

\node[anchor=south west,above right,yshift=0.1em,inner sep=0] at (axis description cs:0,1) {(a) E1};
\end{axis}

\end{tikzpicture}%
        \phantomcaption\label{fig:spectral-density-e1}%
	\end{subfigure}
    \begin{subfigure}[t]{0.495\textwidth}
        \vskip 0pt
		\begin{tikzpicture}

\definecolor{darkgray176}{RGB}{176,176,176}
\definecolor{lightgray204}{RGB}{204,204,204}
\definecolor{darkorange25512714}{RGB}{255,127,14}

\begin{axis}[
axis on top,
colorbar,
colorbar style={
	ylabel={PSD},
	width=0.05\figurewidth,
	at={(1.1,0)},
	anchor=south west,
	ytick={0,0.004,...,0.014},
	every y tick scale label/.style={
		at={(yticklabel* cs:1.2,0)},
		anchor=near yticklabel
	},
},
point meta min=0,
point meta max=0.014,
colormap/viridis,
height=\figureheight,
tick align=outside,
tick pos=left,
width=\figurewidth,
x grid style={darkgray176},
xlabel={$T$\,/\,s},
xlabel style={yshift=0.25em},
xmin=5, xmax=70,
xtick style={color=black},
y grid style={darkgray176},
ylabel={$\lambda$\,/\,m},
scaled y ticks=false,
yticklabel style={
	/pgf/number format/fixed,
	/pgf/number format/precision=3
},
ymin=0.05, ymax=0.2,
ytick style={color=black},
xtick={5,25,...,70},
clip mode=individual,
legend cell align={left},
legend style={
    legend columns = 2,
	fill opacity=0.8,
	draw opacity=1,
	text opacity=1,
	at={(0.999,0.999)},
	anchor=north east,
	draw=lightgray204,
    font=\scriptsize,
    legend image post style={scale=0.55}
},
]
\addlegendimage{line legend,darkorange25512714,very thick}
\addlegendentry{Simulation}

\addplot graphics[xmin=5,xmax=70,ymin=0.05,ymax=0.2]{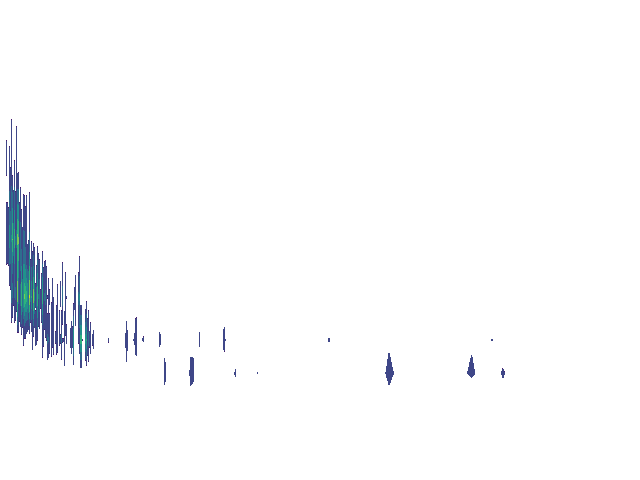};
\addplot graphics[xmin=5,xmax=70,ymin=0.05,ymax=0.2]{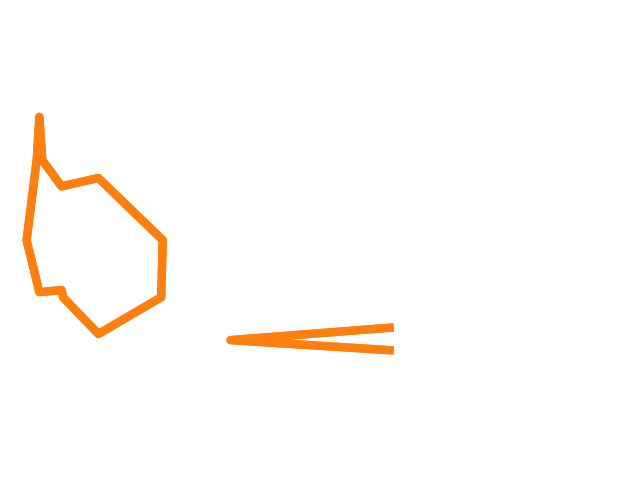};

\node[anchor=south west,above right,yshift=0.1em,inner sep=0] at (axis description cs:0,1) {(b) E4};
\end{axis}

\end{tikzpicture}%
        \phantomcaption\label{fig:spectral-density-e4}%
	\end{subfigure}
	\caption{\label{fig:spectral-density}
        Power spectral density (PSD) of wavelength $\lambda$ and period $T$ for \subref{fig:spectral-density-e1} E1  and \subref{fig:spectral-density-e4} E4  from \citet{pascal2021VariabilityAntiduneMorphodynamics}.
        The orange line is the outermost hull of the simulation results' PSD.
	}
\end{figure}

To obtain a more precise comparison of the numerical and experimental data in terms of their bedform size and fluctuations, we have followed \cite{pascal2021VariabilityAntiduneMorphodynamics} and computed  the power spectral density (PSD) from the square of the two-dimensional discrete Fourier transform of $h_{\text{b}}(x,t)$ with respect to the streamwise position $x$ and time $t$.
We have normalized the PSD by the total number of samples available for $x$ and $t$.
The results are shown in \cref{fig:spectral-density}, along with the experimental PSD.
Although there is not a perfect match between the spectra for the experimental and numerical bedforms, the range of wavelengths and periods agree reasonably well.
For the numerical data, the range of periods and wavelengths in the spectra is slightly narrower than in the experiments, which can be attributed to the much shorter data series from the simulations ($45$\,s simulation time versus more than $4600$\,s in E1, and $2800$\,s in E4 in the experiments).
\par

\begin{figure}
	\centering
	\setlength{\figureheight}{0.3\textwidth}
	\setlength{\figurewidth}{0.45\textwidth}
	\begin{subfigure}[t]{0.495\textwidth}
        \vskip 0pt
		\begin{tikzpicture}

\definecolor{darkgray176}{RGB}{176,176,176}
\definecolor{lightgray204}{RGB}{204,204,204}
\definecolor{darkorange25512714}{RGB}{255,127,14}
\definecolor{steelblue31119180}{RGB}{31,119,180}

\begin{axis}[
axis on top,
colorbar style={
	ylabel={PSD},
	width=0.05\figurewidth,
	at={(1.1,0)},
	anchor=south west,
	ytick={0,0.008,...,0.016},
	every y tick scale label/.style={
		at={(yticklabel* cs:1.2,0)},
		anchor=near yticklabel
	},
},
point meta min=0,
point meta max=0.016,
colormap/viridis,
height=\figureheight,
tick align=outside,
tick pos=left,
width=\figurewidth,
x grid style={darkgray176},
xlabel={$\lambda$\,/\,m},
xlabel style={yshift=0.25em},
xmin=0.05, xmax=0.2,
xtick style={color=black},
y grid style={darkgray176},
ylabel={$c$\,/\,(m/s)},
scaled y ticks=false,
yticklabel style={
	/pgf/number format/fixed,
	/pgf/number format/precision=3
},
xticklabel style={
	/pgf/number format/fixed,
	/pgf/number format/precision=2
},
ymin=0, ymax=0.015,
ytick style={color=black},
xtick={0.05,0.1,0.15,0.2},
clip mode=individual,
legend cell align={left},
legend style={
    legend columns = 1,
	fill opacity=0.8,
	draw opacity=1,
	text opacity=1,
	at={(0.999,0.001)},
	anchor=south east,
	draw=lightgray204,
    font=\scriptsize,
    legend image post style={scale=0.55}
},
]
\addlegendimage{line legend,steelblue31119180,very thick}
\addlegendentry{Experiment}
\addlegendimage{line legend,darkorange25512714,very thick}
\addlegendentry{Simulation}

\addplot graphics[xmin=0.05,xmax=0.2,ymin=0,ymax=0.015]{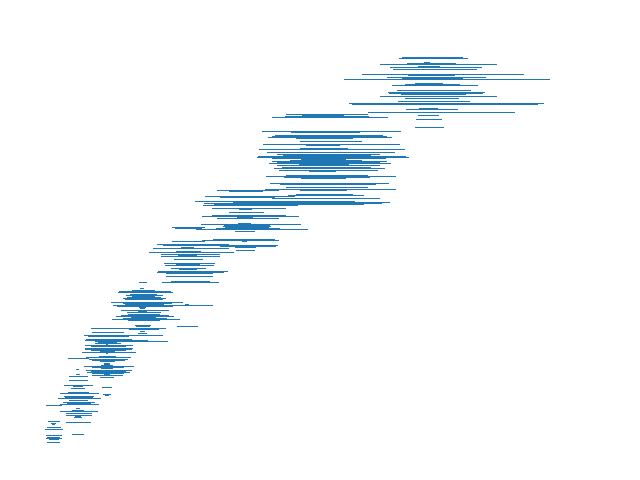};
\addplot graphics[xmin=0.05,xmax=0.2,ymin=0,ymax=0.015]{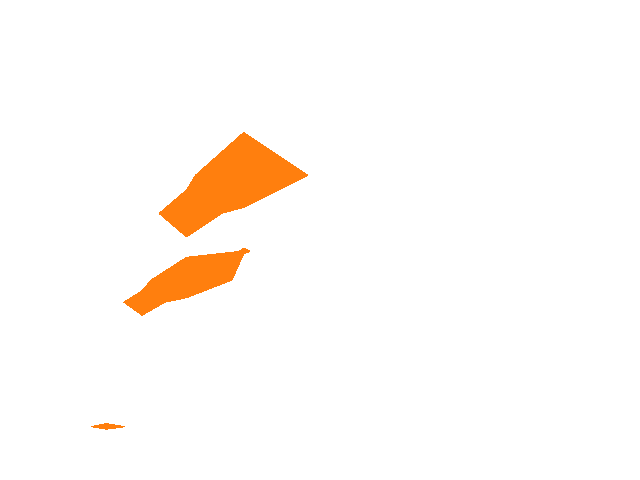};

\node[anchor=south west,above right,yshift=0.1em,inner sep=0] at (axis description cs:0,1) {(a) E1};
\end{axis}

\end{tikzpicture}%
        \phantomcaption\label{fig:celerity-e1}%
	\end{subfigure}
    \begin{subfigure}[t]{0.495\textwidth}
        \vskip 0pt
		\begin{tikzpicture}

\definecolor{darkgray176}{RGB}{176,176,176}
\definecolor{lightgray204}{RGB}{204,204,204}
\definecolor{darkorange25512714}{RGB}{255,127,14}
\definecolor{steelblue31119180}{RGB}{31,119,180}

\begin{axis}[
axis on top,
colorbar style={
	ylabel={PSD},
	width=0.05\figurewidth,
	at={(1.1,0)},
	anchor=south west,
	ytick={0,0.007,...,0.014},
	every y tick scale label/.style={
		at={(yticklabel* cs:1.2,0)},
		anchor=near yticklabel
	},
},
point meta min=0,
point meta max=0.014,
colormap/viridis,
height=\figureheight,
tick align=outside,
tick pos=left,
width=\figurewidth,
x grid style={darkgray176},
xlabel={$\lambda$\,/\,m},
xlabel style={yshift=0.25em},
xmin=0.05, xmax=0.2,
xtick style={color=black},
y grid style={darkgray176},
ylabel={$c$\,/\,(m/s)},
scaled y ticks=false,
yticklabel style={
	/pgf/number format/fixed,
	/pgf/number format/precision=3
},
xticklabel style={
	/pgf/number format/fixed,
	/pgf/number format/precision=2
},
ymin=0, ymax=0.025,
ytick style={color=black},
xtick={0.05,0.1,0.15,0.2},
clip mode=individual,
legend cell align={left},
legend style={
    legend columns = 1,
	fill opacity=0.8,
	draw opacity=1,
	text opacity=1,
	at={(0.999,0.001)},
	anchor=south east,
	draw=lightgray204,
    font=\scriptsize,
    legend image post style={scale=0.55}
},
]
\addlegendimage{line legend,steelblue31119180,very thick}
\addlegendentry{Experiment}
\addlegendimage{line legend,darkorange25512714,very thick}
\addlegendentry{Simulation}

\addplot graphics[xmin=0.05,xmax=0.2,ymin=0,ymax=0.025]{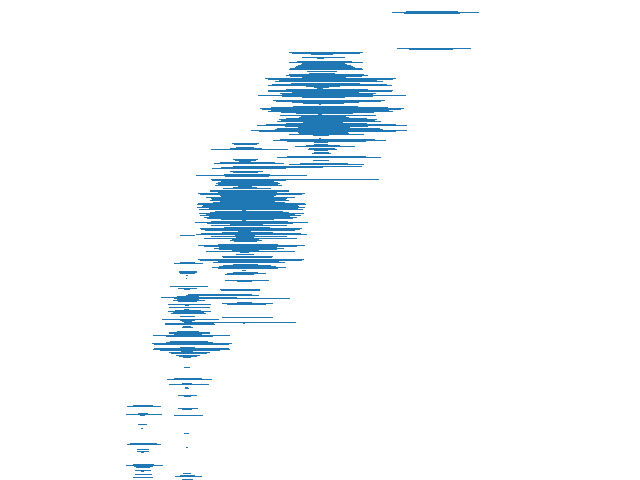};
\addplot graphics[xmin=0.05,xmax=0.2,ymin=0,ymax=0.025]{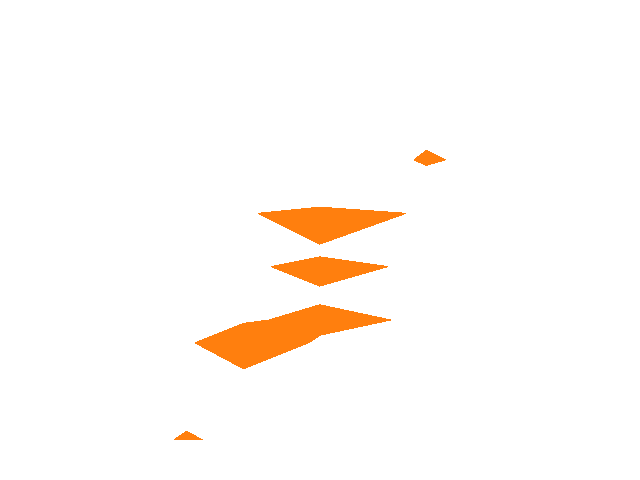};

\node[anchor=south west,above right,yshift=0.1em,inner sep=0] at (axis description cs:0,1) {(b) E4};
\end{axis}

\end{tikzpicture}%
        \phantomcaption\label{fig:celerity-e4}%
	\end{subfigure}
	\caption{\label{fig:celerity}
        Power spectral density contour of celerity $c$ and wavelength $\lambda$ for \subref{fig:celerity-e1} E1  and \subref{fig:celerity-e4} E4  from \citet{pascal2021VariabilityAntiduneMorphodynamics}.
	}
\end{figure}

We compared experimental and simulated results further by obtaining the spectra in the celerity--wavelength domain (\cref{fig:celerity}).
The celerity $c = \lambda/T$, that is, the movement speed of the dunes, was computed as the ratio between wavelength $\lambda$ and period $T$ from the spectra in \cref{fig:spectral-density}.
Similarly to the PSD in the $\lambda-T$ plane, a fairly good agreement was found between the simulations and the experiments. More precisely, the celerities in the simulations range from $2$ to $12$\,mm/s in E1 (experiment: $\approx 2-15$\,mm/s) and from $2$ to $18$\,mm/s in E4 (experiment: $\approx 5-30$\,mm/s).
It is important to note that the simulations reproduced the trends observed experimentally, that for a given wavelength, celerity increases with bedload transport rate, while for a constant bedload transport rate, celerity increases with increasing antidune wavelength.
This latter trend is opposite to that commonly observed for morphodynamics in subcritical flows \citep[e.g.,][]{coleman2011}.
\par

\begin{table}
  \centering
  \begin{tabular}{
                    >{\raggedright}m{0.075\textwidth}
		    	    >{\centering\arraybackslash}m{0.075\textwidth}
                    >{\centering\arraybackslash}m{0.075\textwidth}
			        >{\centering\arraybackslash}m{0.075\textwidth}
			        >{\centering\arraybackslash}m{0.075\textwidth}
			        >{\centering\arraybackslash}m{0.075\textwidth}
			        >{\centering\arraybackslash}m{0.075\textwidth}
			        >{\centering\arraybackslash}m{0.1\textwidth}
			        >{\centering\arraybackslash}m{0.1\textwidth}
                    >{\centering\arraybackslash}m{0.1\textwidth}}
    Label & $h_{0}/d_{\text{50}}$ & $h_{0}/d_{\text{50}}$ & $\bar{h}/d_{\text{50}}$ & $\text{tan}\psi$ & $\text{tan}\psi$ & $\Theta$ & $\Theta$ & $q_{\text{b}}\ast$ & $q_{\text{b}}\ast$\\
    & exp. & sim. & sim. & exp. & sim. & exp. & sim. & exp. & sim.
    \\
    \midrule
    E1 & $2.86$ & $2.97$ & $3.02$ & $0.051$ & $0.044$ & $0.085$ & $0.086$ & $0.033$ & $0.020$ \\
    E4 & $3.59$ & $3.59$ & $3.10$ & $0.052$ & $0.051$ & $0.108$ & $0.119$ & $0.100$ & $0.052$\\
  \end{tabular}
    
  \caption{\label{tab:variable-summary}
    Summary of experimental (exp.) and simulated (sim.) variables, where $\psi$ is the mean bed slope, $\Theta=R_{\text{b}}\text{tan}\psi/[d_{\text{50}}(\rho_{\text{p}}/\rho_1 - 1)]$ is the  Shields number, and $q_{\text{b}}\ast=q_{\text{b}}/[(\rho_{\text{p}}/\rho_1 - 1)gd_{\text{50}}^{\text{3}}]^{\text{1/2}}$ is the Einstein bed load number.
  }
\end{table}

\subsection{Hydraulic variables and sediment transport}\label{subsec:bedload}

To further compare the experimental and simulated flow and sediment transport conditions in a quantitative sense, the normalized average water depth, the bottom slope, the Shields number and the Einstein bed load number are defined and summarized in \cref{tab:variable-summary}.
It must be noted that the bottom profile in the simulations was intentionally horizontal, and thus the slope $\psi$ considered in \cref{tab:variable-summary} is the tangent to the acting force imposed in $x$--$z$ direction.
Assuming a steady-uniform flow where the fluid gravitational force is balanced by boundary friction, the average bed shear stress was computed with the depth-slope product $\rho_1 g R_{\text{b}}\text{tan}\psi$. Here, $R_\text{b}$ is the hydraulic radius.
\par

\begin{figure}
	\centering
	\setlength{\figureheight}{0.3\textwidth}
	\setlength{\figurewidth}{\textwidth}
	\input{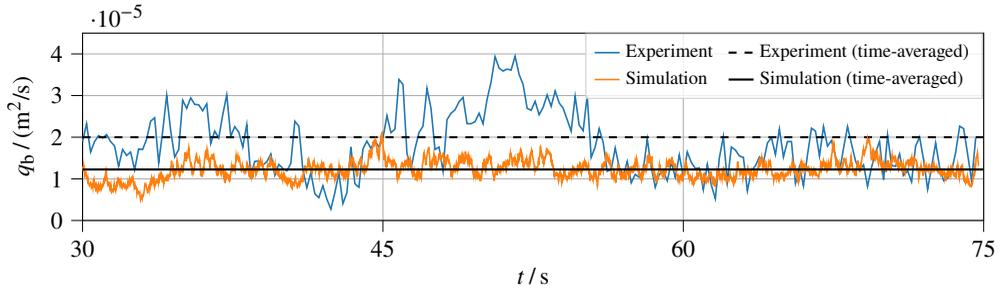}%
	\caption{\label{fig:bedload-rate-e1}
	    Bedload transport rate $q_{\text{b}}(t)$ as measured in the experiment E1 from \citet{pascal2021VariabilityAntiduneMorphodynamics} and in the numerical simulations.
        Time-averaged values were computed for a time range of $2804$\,s and $45$\,s for the experiment and simulation, respectively.
	}
\end{figure}

Comparing the numbers of \cref{tab:variable-summary}, the numerical and experimental Shields numbers are very similar.
We note that in our simulations the horizontal boundary conditions were periodic, that is, the domain was not bound by lateral walls as in the physical experiments.
Therefore, for the simulations, the boundary shear stress in the Shields number was computed based on the average water depth ($R_{\text{b}}=\bar{h}$) from $t=30$\,s to $t=75$\,s; whereas, for the experiments, the boundary shear stress considered the hydraulic radius as a correction for the effect of the lateral wall roughness.
The simulated sediment transport rates were lower than the experimental mean values by a factor of about $2$.
Nevertheless, the increase in sediment transport rate from case E1 to E4 was similar in the simulations and the experiments (roughly an increment by a factor $3$).
\Cref{fig:bedload-rate-e1} shows the evolution of the simulated sediment transport rates with time for E1, together with a representative time window of similar duration of the experiments.
Compared to the simulation data, the experimental transport rates show larger fluctuations.
These variations are likely due to sampling, which captured all particles ejected from the downstream end of the channel in the experiment, as opposed to the averaged sediment transport rates over the entire simulation domain.
It is, therefore, remarkable that the simulated and experimental transport rates match well in the last $15$\,s of the time series.
Some reasons for the underestimation of the average sediment transport rates in the simulations can be related to the shape of the particles 
\citep[natural gravel in the experiments, spheres in the simulations, cf.\ ][]{deal2023grain}, to experimental uncertainties as for instance, additional moment added by the sediment feeding system, and to the strong non-linear dependency of sediment transport rates on boundary conditions and the wall shear stresses.
Notwithstanding, the comparison between numerical and experimental hydraulic variables and sediment transport rates must be rated as satisfactory, and given the good correspondence between the simulations and the experiments, we conclude that our simulation approach is adequate to investigate UMAs with high fidelity.
\par

\section{Conclusion}

We have performed particle-resolved numerical simulations of supercritical turbulent flows ($\mathit{Fr}>1$) over an erodible granular bed of spherical particles.
Our goal was to numerically reproduce the experimental work of \citet{pascal2021VariabilityAntiduneMorphodynamics}, in which upstream migrating antidunes developed on a bed of natural gravel sheared by a shallow-water flow (water depth $\approx3$ times the sediment grain size).
Supercritical shallow flows over erodible beds are extremely unstable due to the strong feedback between the dynamics of the bottom and the wavy free surface.
Our results demonstrate that one can accurately reproduce the bedform dynamics at different supercritical flow conditions by using a massively parallel simulation approach relying on fully coupled fluid--particle--gas interactions.
In our simulations, no suspended load was present and the bed instability was naturally excited by the simulated flow.
Hence, the range of amplitudes, wavelengths, and celerities of the self-generated upstream migrating bedforms were results of the simulations and agree well with the experimental patterns reported by \citet{pascal2021VariabilityAntiduneMorphodynamics}.
Particularly, the averaged bed Shields stress from the simulations as a key quantity to predict sediment transport rates matched very well with the experimental values.
\par

The vast amount of data generated by simulations in a non-intrusive manner such as those presented in this work can be of great use to supplement experimental measurements under challenging supercritical flow conditions.
The detailed data of particle and fluid motion available through our simulation approach opens a multitude of possibilities.
The present study, therefore, encourages further simulation campaigns to understand antidune mechanics and the physical controls on bedform initiation and morphodynamics in supercritical flows.
\par

\vspace{\baselineskip}

\noindent{\bf Supplementary data\bf{.}} \label{SupMat} Animations of the simulations are available online.\\
\noindent{\bf Acknowledgements\bf{.}}
The authors gratefully acknowledge the Gauss Centre for Supercomputing e.V.\ (www.gauss-centre.eu) for funding this project by providing computing time on the GCS Supercomputer SuperMUC at Leibniz Supercomputing Centre (www.lrz.de).\\
\noindent{\bf Funding\bf{.}}
The authors thank the Deutsche Forschungsgemeinschaft (DFG, German Research Foundation) for funding the projects 408062554, 434946896, 433735254, and 428445330.
This work was supported by the SCALABLE project.
This project has received funding from the European High-Performance Computing Joint Undertaking (JU) under grant agreement No 956000.
The JU receives support from the European Union’s Horizon 2020 research and innovation programme and France, Germany, the Czech Republic.\\
\noindent{\bf Declaration of Interests\bf{.}}
The authors report no conflict of interest.\\
\noindent{\bf Author ORCID\bf{.}}
C.\ Schwarzmeier, \url{https://orcid.org/0000-0002-5756-3985};
C.\ Rettinger, \url{https://orcid.org/0000-0002-0605-3731};
S.\ Kemmler, \url{https://orcid.org/0000-0002-9631-7349};
J.\ Plewinski, \url{https://orcid.org/0000-0002-4815-7643};
F.\ N\'{u}\~{n}ez-Gonz\'{a}lez, \url{https://orcid.org/0000-0002-3676-2715};
H.\ Köstler, \url{https://orcid.org/0000-0002-6992-2690};
U.\ Rüde, \url{https://orcid.org/0000-0001-8796-8599};
B.\ Vowinckel, \url{https://orcid.org/0000-0001-6853-7750}\\
 \noindent{\bf Author contributions\bf{.}}
 C.\ Schwarzmeier: Conceptualization, Methodology, Software, Formal Analysis, Validation, Investigation, Data Curation, Writing -- Original Draft, Writing -- Review \& Editing, Visualization, Supervision, Project Administration;
 C.\ Rettinger: Conceptualization, Methodology, Software, Formal Analysis, Validation, Investigation, Data Curation, Writing -- Original Draft, Visualization, Supervision, Project Administration;
 S.\ Kemmler: Software, Validation, Investigation, Writing -- Original Draft;
 J.\ Plewinski: Software, Validation, Writing -- Original Draft;
 F.\ N\'{u}\~{n}ez-Gonz\'{a}lez: Conceptualization, Writing -- Original Draft, Writing -- Review \& Editing;
 H.\ Köstler: Resources, Funding acquisition;
 U.\ Rüde: Resources, Funding acquisition, Writing -- Original Draft; 
 B.\ Vowinckel: Conceptualization, Writing -- Original Draft, Writing -- Review \& Editing;


\bibliography{library.bib}{}
\bibliographystyle{jfm}

\end{document}